\documentclass{llncs}
\usepackage{hyperref}
\usepackage{ifpdf}
\ifpdf
 \usepackage[pdftex]{graphicx}
\else
 \usepackage[dvips]{graphicx}
\fi
\usepackage{color}
\usepackage{url}
\usepackage{svn-multi}

\newcommand{\webgeometrylab}{{\em Web Geometry Laboratory}}
\newcommand{\WGL}{{\em WGL}}

\title{A Web Environment for Geometry} 

\author{Pedro Quaresma\inst{1,2} 
  \and Vanda Santos\inst{2,3}
  \and Milena Mari{\'c}\inst{4} }

\institute{
  Department of Mathematics, University of Coimbra, Portugal
  \and
  CISUC, Coimbra, Portugal 
  \and   
  University National of Timor Lorosa'e, East-Timor
  \and 
  Faculty of Mathematics, University of Belgrade, Serbia\\
  \email{pedro@mat.uc.pt}, \email{vsantos7@gmail.com}, \email{milena.maric.f@gmail.com}   
}

\date{}

\begin{document}

\maketitle

\begin{abstract}
  The Web Geometry Laboratory, {\WGL}, is a blended-learning,
  collaborative and adaptive, Web environment for geometry. It
  integrates a well known dynamic geometry system.

  In a collaborative session, exchange of geometrical and textual
  information between the user engaged in the session is possible.

  In a normal work session (stand-alone mode), all the geometric steps
  done by the students are recorded, alongside the navigation
  information, allowing, in a latter stage, their teachers to ``play
  back'' the students sessions, using that info to assert the students
  level and adjust the teaching strategies to each individual student.

  Teachers can register and begin using one of the public servers,
  defining students, preparing materials to be released to the
  students, open collaborative sessions, etc.

  Using an action research methodology the {\WGL} system is being
  developed, validated through case-studies, and further improved, in
  a cycle where the implementation steps are intertwined with case
  studies.

  \keywords{adaptive learning, collaborative learning,
    blended-learning, dynamic geometry} 
\end{abstract}

\section{The Web Geometry System}
\label{sec:wgl}

The \webgeometrylab\ (v1.4) is a Web client/server application; the
server must be hosted by a Web-server (e.g. Apache server) the clients
may use any Web-browser available. The database (to keep:
constructions; users information, constructions permissions, etc.);
the dynamic geometry system (DGS), JavaScript applet; the synchronous
and asynchronous interaction, are all implemented using free
cross-platform software, namely GeoGebra, PHP, JavaScript, AJAX, JSON,
JQuery and MySQL. Also Web-standards like HTML5, CSS style-sheets and
XML. The \WGL\ is an internationalised system with the English
language as the default language and already localised to the
Portuguese and Serbian languages. It is an open-source
system\footnote{\url{http://webgeometrylab.sourceforge.net/}},
versions of the server are available to be installed on Linux systems
(or other systems through virtual machines).

The last version of WGL (1.4) introduces a total separation between
development branches: stable; testing; unstable (development), given a
added stability to the public (stable) server and allowing a public
availability of the code. Apart many small improvements the major new
features are: the text chat; the exchange of geometric information
between the group and individual windows and the saving of the
students work to their own scrapbook, when in a collaborative
sessions; the ``record \& play'' of student's sessions, i.e. the
adaptive module (at a prototype stage in~\cite{Quaresma2014b}); the
JavaScript/HTML5 DGS applet (instead of the Java applet).

Two instances of the \WGL\ server are available, one in
Portugal,\footnote{\url{http://hilbert.mat.uc.pt/WebGeometryLab/}}
another in Serbia.\footnote{\url{http://jason.matf.bg.ac.rs/wgl/}}
Users can log on to the system using the anonymous student-level user,
but without access to collaborative sessions. For more advanced use, a
user must register and then be confirmed by the administrator. During
the last three years these systems have been intensively used, e.g.,
for testing collaborative learning in teaching
geometryr~\cite{Maric2013,Quaresma2013,Quaresma2014b,Santos2013c}. Please
feel free to contact the authors if you want to use the \WGL\
platform, accessing the platform as a teacher.

In any {\WGL} server there are four distinct types of users:
administrators, teachers, students and anonymous visitors. The
administrator(s) main role is the administration of teachers. They
have also access to the log-in information off all users, information
that can be used to streamline the server. 

The teachers are privileged users in the sense that they will be
capable of defining other users; their students. In the beginning of
each school year the teachers should define all their classes, the
students in each class and, if needed, the aggregation of the students
into groups

The students, each linked to a given teacher, are able to work in the
platform, performing tasks created by their teachers and/or pursuing
their own work. The students are unable to create other users.

Finally, the anonymous visitor is a student-type user, not linked to
any teacher and because of that, unable to participate in
collaborative sessions. The purpose of this type of user is solely to
allow unregistered users to test the \WGL\ platform.

There are two distinct modes for the students to interact with the
{\WGL} system. The collaborative sessions and the regular
(stand-alone) sessions. These two distinct modes are controlled by the
teachers. In a collaborative session the students, working in groups,
have some specific assignment to fulfil and they will do it in a
collaborative way, exchanging geometric and textual information to
reach the common goal. In a regular session the students will be
working alone, they can share constructions with the other users of
the platform but all this exchange of information will be
asynchronous.

\section{The Collaborative Module} 
\label{sec:collaborativemodule}

Planning a collaborative working session the teacher has to decide how
to group the students and the design of the tasks to be solved
collaboratively, i.e., prepare a set of geometric constructions,
starting points for tasks to be completed during the class;
illustrative cases; etc.

In a \WGL\ collaborative session the students will solve the tasks
proposed by their teachers, being able to exchange geometric and
textual information, producing the geometric constructions in a
collaborative fashion.

The students engaged in a collaborative session will always be in
working groups, with access to the material prepared by the teacher
and with access to two DGS applets. One of those DGS applets is for
their own work, the other is where the group construction is being
done. The {\em group-construction} is shared by all the members of a
given group, one of the students will have the lock over the
construction, all the other group members will see the work being done
(synchronised every 20s). At any given moment the student can release
the lock, which can be claimed by any student in the group.

At the same time, the students has their own work-space, this can be
used to: follow the work that is being done by the group
representative; develop their own constructions; to anticipate the
group construction; to develop auxiliary constructions. In this
work-space the saving of the work being done is the responsibility of
the student.

The students have the possibility of exchange constructions between
DGS work\-space windows. The students without the lock should be able
to ``import'' the group construction to his/her own work-space. The
student with the lock adds to that, the possibility of exporting the
construction to the group work-space. A chat is provided to allow the
exchange of short messages between all the members of the group,
including the teacher.

Apart from being responsible for setting the collaborative session and
being able to assess its results at the end, the teacher has also
access to a DGS work-space window where he/she can follow the work of
all the groups and all the individual students in each group.

\section{The Adaptive Module} 
\label{sec:adaptivemodule}

To be able to build individual student's profiles and/or individual
learning paths, the system collects information about the student's
interactions when in the stand-alone mode, i.e., in a regular work
session.

The system records navigation and also geometric information for each
student. The navigation information is a plain list of all the pages
visited with enter and exit time-stamps. The geometric information is
recorded when the student is using the DGS applet, using JavaScript
listeners of the DGS application programming interface. We record
every step done by the students.

At a later stage the student's teacher is able to see the work done by
the student, play step by step, play in a regular speed, play in a
fast forward fashion. In this way the teacher can analyse the path
used by the students to solve a given task, getting information that
can be used to assert the student's van Hiele
level~\cite{Crowley1987}.

\section{Access to the System}
\label{sec:accessWGL}

The \WGL\ public servers can be used by any interested teacher. The
International/Portuguese server is
\url{http://hilbert.mat.uc.pt/WebGeometryLab}, the Serbian server is
\url{http://jason.matf.bg.ac.rs/wgl}. After registration (subject to
validation) a teacher can create classes and use the system as a
geometry laboratory or as platform for homework tasks. In a
stand-alone fashion or in collaborative sessions.

We performed two different set of studies, one in Portugal, in
classroom mode, and another in Serbia, in remote access
mode\footnote{\emph{Web Geometry Laboratory: Case Studies in Portugal
    and Serbia}, submitted to \emph{Educational Technology Research
    and Development}, May 2015}. The first set of studies was done
using {\WGL} version 1.2, 
still without the group-wise communication channel (chat). The second
set of studies were done using {\WGL} version 1.3, already with the
chat communication channel, among other developments done in the
platform. The platform was positively received by teachers and
students, improving their learning experience~\cite{Santos2013c}. The
case-studies were/are used to improve the system but also to publicise
the system, training teachers in its use.

A forum (phpBB forum) is provided to allow the exchange of information
between users.

\section{Conclusions and Future Work}
\label{sec:ConclusionsFutureWork}

\paragraph{Related Systems}\label{sec:Related Systems} There are
several DGS available (see~\cite{wikipediaListDGSs} for a
comprehensive list) but none of them defines an environment where the
DGS is integrated into a learning platform with collaborative and
adaptive features. In~\cite{Quaresma06d,Quaresma2007,Santos08} we can
find accounts of DGSs and geometric automated theorem provers (GATPs)
integration and the integration of those tools in learning
environments but always partial integrations not building any kind of
collaborative, adaptive blended-learning platform.  Some learning
environments in the area of geometry have been developed,
e.g. Tabulae~\cite{Moraes2005} and GeoThink~\cite{Moriyon2008}. The
{\WGL} distinguishes itself relying on an external DGS, allowing in
this way to possess a full fledged DGS, well known by its users and
supported by its developers. The well grounded permissions system and
the capability that this opens for a personalised contact with the
platform, is also something in favour of {\WGL}. The many case-studies
already conducted, validating the {\WGL} goals, and the
internationalisation, i.e. the ability to receive translations into
different languages (Tabul\ae\ lacks this feature), are also positive
points for {\WGL}.

\paragraph{Conclusions and Future Work}\label{sec:futurework}

At the moment the adaptive module only collects the student's
information and allow the teachers to ``play'' that information. A
first step ahead, already planned, will give the teachers the
possibility of building students profiles or individualised learning
paths. A second, more ambitious, step would give the system some
capabilities of automatic construction of those profiles and/or
learning paths.

A second development planned is the integration of a GATP. To be able
to provide a formal validation of geometric properties, e.g. ``{\em
  two lines are perpendicular, because \ldots}'' and also to support
the automatic or semi-automatic adaptive features, e.g. one-step
guidance, formal reasoning and visual proofs.

The Web Geometry Laboratory is a blended-learning, collaborative,
adaptive, Web environment for geometry already being used by teachers
in Portugal and Serbia and we expect that its user base can grown not
only in those countries but also in other countries.

\section*{Acknowledgments}

The first author is partially supported by the iCIS project
(CENTRO-07-ST24-FEDER-002003), co-financed by QREN, in the scope of
the Mais Centro Program and European Union's FEDER.

\bibliographystyle{plain}

\newcommand{\noopsort}[1]{} \newcommand{\singleletter}[1]{#1}

\end{document}